\documentclass[]{tMPH2e}
\usepackage{rotating}
\usepackage[matrix,frame,arrow]{xy}
\newcommand{\qw}[1][-1]{\ar @{-} [0,#1]}
\newcommand{\qwx}[1][-1]{\ar @{-} [#1,0]}
\newcommand{\qwxdt}[1][-1]{\ar @{--} [#1,0]}
\newcommand{\cw}[1][-1]{\ar @{=} [0,#1]}

\newcommand{\gate}[1]{*{\xy *+<.6em>{#1};p\save+LU;+RU **\dir{-}\restore\save+RU;+RD **\dir{-}\restore\save+RD;+LD **\dir{-}\restore\POS+LD;+LU **\dir{-}\endxy} \qw}
\newcommand{\meter}{\gate{\xy *!<0em,1.1em>h\cir<1.1em>{ur_dr},!U-<0em,.4em>;p+<.5em,.9em> **h\dir{-} \POS <-.6em,.4em> *{},<.6em,-.4em> *{} \endxy}}

\newcommand{\control}{*-=-{\bullet}}

\newcommand{\ctrl}[1]{\control \qwx[#1] \qw}
\newcommand{\ctrldt}[1]{\control \qwxdt[#1] \qw}

\newcommand{\targ}{*{\xy{<0em,0em>*{} \ar @{ - } +<.4em,0em> \ar @{ - } -<.4em,0em> \ar @{ - } +<0em,.4em> \ar @{ - } -<0em,.4em>},*+<.8em>\frm{o}\endxy} \qw}
\newcommand{\rstick}[1]{*!L!<-.5em,0em>=<0em>{#1}}
\newcommand{\lstick}[1]{*!R!<.5em,0em>=<0em>{#1}}

\newcommand{\Qcircuit}{\xymatrix @*=<0em>}

\newcommand{\bra}[1]{\langle#1|}
\newcommand{\ket}[1]{|#1\rangle}
\newcommand{\ad}{a^\dagger}
\newcommand{\bfx}{\mathbf{x}}
\newcommand{\TwoE}[4]{\ad_{#1}\ad_{#2}a_{#3}a_{#4}}
\newcommand{\I}{\mathbf{1}}
\newcommand{\X}{\sigma^x}
\newcommand{\Y}{\sigma^y}
\newcommand{\Z}{\sigma^z}

\begin{document}
\doi{10.1080/0026897YYxxxxxxxx}
 \issn{1362–3028}
\issnp{0026–8976}
\jvol{00}
\jnum{00} \jyear{2010} 

\markboth{Whitfield et al.}{arXiv:1001.3855}

\articletype{}

\title{Simulation of Electronic Structure Hamiltonians Using Quantum Computers}
\author{James D. Whitfield$^{a}$, Jacob Biamonte$^{a,b}$, and Al\'{a}n Aspuru-Guzik$^a$\thanks{Email: whitfield@chemistry.harvard.edu \vskip.5ex Email: jacob.biamonte@qubit.org\vskip.5ex Email: aspuru@chemistry.harvard.edu}\\\vspace{6pt}  $^{a}${\em{Harvard University, Department of Chemistry and Chemical Biology, 12 Oxford St., Cambridge, MA, 02138, USA}};\\$^{b}${\em{Oxford University Computing Laboratory, Oxford OX1 3QD, UK}}}

\maketitle

\begin{abstract}
Over the last century, a large number of physical and mathematical developments paired with rapidly
advancing technology have allowed the field of quantum chemistry to advance dramatically.
However, the lack of computationally efficient methods for the exact simulation of quantum systems on classical computers presents a limitation of current computational approaches. 
We report, in detail, how a set of pre-computed molecular integrals can be used to explicitly create a quantum circuit, i.e. a sequence of elementary quantum operations, that, when run on a quantum computer, to obtain
the energy of a molecular system with fixed nuclear geometry using the quantum phase estimation algorithm.
We extend several known results related to this idea and discuss the adiabatic state preparation procedure for preparing the input states used in the algorithm. With current and near future quantum devices in mind, we provide a complete example using the hydrogen molecule, of how a chemical Hamiltonian can be simulated using a quantum computer.    
\bigskip
\end{abstract}

\section{Introduction}
Theoretical and computational chemistry involves solving the equations of motion that govern quantum systems by analytical and numerical methods~\cite{SO96,HJO00}. Except in standard cases such as, the harmonic oscillator or the hydrogen atom, analytic solutions are not known and computational methods have been developed.  

Although classical computers have tremendously aided our understanding of chemical systems and their processes, the computational cost of the numerical methods for solving Schr\"odinger's equation grows rapidly with increases in the quality of the description.  
Research is ongoing to improve computational methods, but large molecules and large basis sets have remained a consistent problem despite the exponential growth of computational power of classical computers~\cite{Sherrill2010}.


Theoretical computer science suggests that these limitations are not mere shortcomings of the programmers but could stem from the inherent difficultly of simulating quantum systems. Extensions of computer science using quantum mechanics led to the exploitation of the novel effects of quantum mechanics for computational purposes resulting in several proposals for quantum computers~\cite{LJL+10}.  Quantum simulation is the idea of using quantum computational devices for more efficient simulation \cite{Fey82, Llo96}.  Since the dynamics are simulated by a quantum system rather than calculated by a classical system, quantum simulation often offers  exponential advantage over classical simulation for calculation of electronic energies~\cite{AGDL+05}, reaction rates~\cite{LW99,KJLM+08}, correlation functions~\cite{OGKL01} and molecular properties~\cite{KA09}.  Recently, a review of these  techniques and other applications of quantum computing to chemistry has appeared~\cite{Kassal2010}.

This paper does not consider the effect of errors, however it is an important consideration that needs to be taken into account. Quantum error correction methods have been developed to counteract the unwanted effect of quantum noise, however fault tolerant constructions require redundant qubits and only allow a discrete set of gates to be used~\cite{NC01,Got09}.  This is not a serious cause for concern as the conversion from a continuous set of gates to a discrete set of gates only requires a poly-logarithmic overhead~\cite{Kit97}.  Clark et al.~\cite{CBMG08} estimated the resources required to compute the ground state of a one dimensional transverse Ising model and found, using experimental parameters from a proposed ion trap quantum computing implementation, that the fault tolerant constructions would be too costly for straight-forward applications to simulation.  This suggests that quantum simulation without quantum error correction is more feasible for the near future.

The state of the art in experimental realizations of quantum simulation for chemistry is represented by calculations of the energy spectrum of molecular hydrogen first by Lanyon et al.~\cite{LW+09} using an optical quantum computer. Very soon after, Du et al.~\cite{DXP+09} used NMR technology to demonstrate the adiabatic state preparation procedure suggested by Ref.~\cite{AGDL+05} as well as reproduce the energy to higher accuracy.  

The key limitation of both experimental algorithms was the representation of the simulated system's propagator.  Both experiments, relied on the low dimensionality  of the propagator for the minimal basis H$_2$ model considered.  The unitary propagator for a two-level system can be decomposed using the real angles $\alpha,\beta,\gamma$:
\[U=e^{i\alpha}R_{y}(\beta)R_{z}(\gamma)R_{y}(-\beta)\]
Due to this decomposition, longer propagation times corresponding to higher powers  of unitary evolution operator, $U^j$, can be achieved by changing $\alpha$ to $j\alpha$ and $\gamma$ to $j\gamma$, thereby avoiding the need of further decomposition of the unitary operator.  Beyond the two dimensional case, this decomposition is not available.

The objective of this paper is to provide a general decomposition for electronic Hamiltonians and demonstrate this method with an explicit quantum circuit for a single Trotter time step of the minimal basis hydrogen molecule.    This is an extension of the supplementary material from Lanyon et al.~\cite{LW+09}.  

\section{Overview of the quantum algorithm}
The use of the Fourier transform of correlation functions in computational 
chemistry allows for the extraction of information about many properties such
as transport coefficients~\cite{Zwanzig1965} and molecular spectra~\cite{Heller1981}. For the application of quantum
computation to molecular electronic energy calculations, the spectral solution
 of the Schr\"odinger equation~\cite{Feit1982} is pursued.  The key idea of the approach is 
that the Fourier transform, $\mathcal{F}\left[f(t)\right]=\int\exp(i\omega t) f(t)dt$ of the autocorrelation of the time 
evolving state, $P(t)=\bra{\psi(0)}\psi(t)\rangle$, has peaks at the eigenenergies. Typically, the
semiclassical propagator is used for examining vibrational structure e.g.~\cite{Ceotto2009}. Since 
quantum computers can simulate the quantum evolution of electrons efficiently, the Fourier 
transform of the electronic autocorrelation function can be calculated directly.  
The measurement technique presented here collapses an input wave function into an
eigenstate and returns the frequency of the autocorrelation of the time-evolving 
eigenstate by employing the spectral method mentioned before to the extraction
 of molecular electronic energies. In this paper, we focus on the extraction 
of eigenvalues given input states using the time evolution of the state.

The construction of the general quantum circuit to simulate the evolution of the molecular system is performed in three steps:
\begin{enumerate}
	\item Write the Hamiltonian as a sum over products of Pauli spin operators acting on different qubits.  This is described in Section~\ref{sec:Ham} and made possible by the Jordan-Wigner transformation.
	\item Convert each of the operators defined in step (1) into unitary gates such that their sequential execution on a quantum {computer} can be made to recover an approximation to the {unitary propagator, $\exp(-iHt)$}.  This is detailed in Section~\ref{sec:UP}.
	\item The phase estimation algorithm, as described in Section~\ref{sec:PEA}, is then used to approximate the {eigenvalue} of an input eigenstate using the quantum Fourier transform of the time domain propagation.  Section~\ref{appx:StPrep} discusses eigenstate preparation. 
\end{enumerate}
To demonstrate these steps, the construction is applied to the example of the hydrogen molecule in Section~\ref{sec:H2}. The key components of the simulation procedure are depicted in Fig.~1. The next section provides a basic review of the fundamental concepts and notations of molecular quantum chemistry for the benefit of quantum information {scientists} and to establish the notation.  A detailed account of electronic structure methods can be found in the monographs~\cite{HJO00,SO96}.
\begin{figure}[tb]
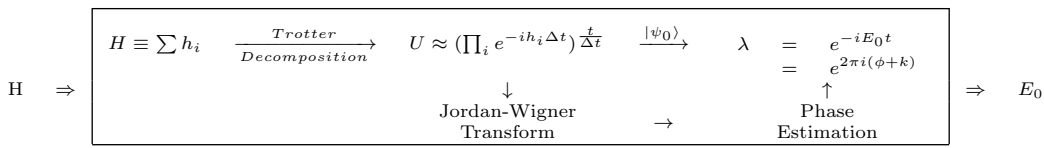

	\scriptsize
\begin{tabular}{cc cc cc cc cc}
    \cline{3-7}
    &&  \multicolumn{1}{|c}{$H\equiv \sum h_i$} &$\xrightarrow[Decomposition]{Trotter}$  & $U\approx(\prod_i e^{-ih_i\Delta t})^{\frac{t}{\Delta t}}$&$\xrightarrow{|\psi_0\rangle}$ & \multicolumn{1}{c|}{\begin{tabular}{rcl}\\[.5ex]$\lambda$&=&$e^{-iE_0 t}$\\&=&$e^{2\pi i (\phi+k)}$\end{tabular}}&&\\
{H}&$\Rightarrow$&      \multicolumn{1}{|c}{ }&&$\downarrow$&&\multicolumn{1}{c|}{$\uparrow$}&$\Rightarrow$&$E_0$\\
&&      \multicolumn{1}{|c}{}   &&\begin{tabular}{c}Jordan-Wigner\\ Transform\end{tabular}&$\rightarrow$&\multicolumn{1}{c|}{\begin{tabular}{c}Phase\\ Estimation\\
\end{tabular}}&&\\
    \cline{3-7}
\end{tabular}
	\caption{
 An algorithmic overview of the steps taken to simulate a chemical Hamiltonian on a quantum computer. The time independent Hamiltonian of a molecular system (to the left of the box) will be decomposed into a sum of Hermitian matrices, ($h_i$) and by means of a Trotter decomposition the unitary propagator $U$ can be constructed.  The Jordan-Wigner mapping  is used to convert the propagator into a sequence of quantum gates.  Phase estimation algorithms are used to evaluate the eigenvalue of a correctly prepared stationary state $\ket{\psi_0}$.
	}
	\label{fig1}
\end{figure}

\section{The Electronic Hamiltonian}\label{sec:Ham}
The Born-Oppenheimer approximation assumes that the wave function of a molecule can be 
expressed as a product of the nuclear wave function and electronic wave function
(parameterized by the nuclear coordinates) due to the difference in mass between electrons and nuclei.   This approximation allows for the solution of the time independent Schr\"odinger equation of the electronic wave function for a given nuclear geometry.

The molecular electronic Hamiltonian\footnote{Throughout this paper, atomic units are used: $\hbar$ ($1.054\cdot 10^{-34}$ J s), the mass of the electron ($9.109\cdot10^{-31}$ kg), the elementary charge ($1.602\cdot 10^{-19}$ C), and the electrostatic force constant ($1/4\pi\epsilon_0=8.988\cdot10^{9}$ N m$^2$ C$^{-2}$) are set to unity.} in second quantized form is given by \cite{SO96,HJO00}:
\begin{equation}\label{eq:ham2nd}
    H=\sum_{p,q}h_{pq}\ad_p a_q +\frac{1}{2}\sum_{p,q,r,s} h_{pqrs} \ad_p \ad_q a_r a_s,
\end{equation}
where the sum is over the single particle basis set described below. The annihilation $\{a_j\}$ and creation operators $\{\ad_j\}$ obey the Fermionic anti-commutation relations (see Table \ref{tbl:2nd_quant}):
\begin{align}
&[a_p, \;a_q]_+=0,& &[a_p, \;\ad_q]_+=\delta_{pq}\I,
\label{:CCR}
\end{align}
where the anti-commutator of operators $A$ and $B$ is defined as $[A,B]_+\equiv AB+BA$ {and $\I$ is the identity operator.}
The annihilation (creation) operators correspond to a set of orbitals, $\{\chi_i\}$,  where each orbital is 
a single-particle wave function composed of a spin and a spatial function, denoted $\sigma_i$ and $\varphi_i$, 
respectively.  As the Hamiltonian considered here commutes with the electron spin operators, $\sigma_i$ is restricted to be one of
two orthogonal functions of a spin variable $\omega$ that we denote $\alpha(\omega)$ and $\beta(\omega)$. Similar Hamiltonians can be found in many physics problems involving Fermionic particles.  

\begin{table}[t]
\tbl{An overview of second quantization for Fermionic particles.  For quantum chemistry, the annihilation and creation operators correspond to removing (adding) an electron into a particular orbital, $\{\chi_k\}$. The anti-symmetry is enforced by the canonical commutation relations, and the $N$-electron wave function is expanded over the configuration state functions of the Fock space. }
{\begin{tabular}{lll}
  \toprule
Second quantization&&\\
\colrule
Creation  operator  & $\ad_i\ket{j_1, \dots, 0_i, \dots, j_n}= \Gamma^\mathbf{j}_i\ket{j_1, ..., 1_i, ..., j_n}$ 
    			& with  $\Gamma^\mathbf{j}_i = \prod^{i-1}_{n=1} (-1)^{j_n}$\\
            & $\ad_i\ket{j_1, \dots, 1_i, \dots, j_n}=0$&\\
Annihilation operator& $a_i\ket{j_1, \dots, 1_i, \dots, j_n}= \Gamma^\mathbf{j}_i\ket{j_1, ..., 0_i, ..., j_n}$& \\
            & $a_i\ket{j_1, \dots, 0_i, \dots, j_n}= 0 $&\\
Canonical commutation relations  & $\{\ad_i, a_j\}=\delta_{ij}\I$&\\
                & $\{a_i, a_j\}=0$ &\\
&&\\
\botrule
\\[.5ex]
\toprule
Fock space&&\\
\colrule
\begin{tabular}{l}Basis vectors,\\ configuration state functions\end{tabular} & \begin{tabular}{lcl}
		$\ket{\mathbf{j}}$&=&$\ket{j_1, j_2, ..., j_N}$\\
		&=&$ \prod^{N}_{p=1} \left(\ad_p\right)^{j_p}\ket{vac}$
		\end{tabular} &where $j_i=0, 1$\\
  Inner product &\multicolumn{2}{l}{$\langle\mathbf{j}\ket{\mathbf{k}}= \prod_{p=1}^N\delta_{j_p, k_p}$} \\

  Vacuum state &{ $\langle vac\ket{vac}=1$}&\\
           & $a_i\ket{vac}=0$&\\
	  Single electron operator, $\hat{A}(x)$	& $\hat{A}_{ij}\ad_ia_j$ &\begin{tabular}[t]{l}where $\hat{A}_{ij}=\int \chi_i(x_1) \hat{A}(x_1) \chi_j(x_1) dx_1$\\ is a single electron operator\\
			{and $\{\chi_k\}$ corresponds to $\{a_k\}$}\end{tabular}\\[1ex]
\botrule
\end{tabular}}
\label{tbl:2nd_quant}
\end{table}

Although any basis can be used, the molecular orbitals are particularly convenient for state preparation reasons discussed below. The molecular orbitals, in turn, are formed as a linear combinations of atomic basis functions~\cite{Fel96,SDES+07}. The coefficients of this expansion are obtained by solving the set of Hartree-Fock equations which arise from the variational minimization of the energy using a single determinant wave function. 
Due to its restriction to a single determinant, the Hartree-Fock solution is a mean field solution and the difference between the Hartree-Fock solution using an infinite basis of atomic orbitals and the exact (correlated) solution defines the electron correlation energy.

The matrix elements $\{h_{pq}\}$ and $\{h_{pqrs}\}$ in Eq.~\eqref{eq:ham2nd} denote the set of one- and two-electron integrals that must be evaluated using a known set of basis functions (the basis set) during the Hartree-Fock procedure.  Ideally, the number of basis functions used would be infinite, but in practice only a finite number of basis functions are used.  By selecting Gaussian single-particle basis functions, these integrals are efficiently computable.  Next, to further establish notation, we develop the explicit form of the integrals $h_{pq}$ and $h_{pqrs}$.

  We denote the set of single-particle spatial functions which constitute the molecular orbitals $\{\varphi_k(x)\}_{k=1}^M$.   Finally, define the set of spin orbitals as $\{\chi_p(\mathbf{x})\}_{p=1}^{2M}$ with $\chi_p=\varphi_i\sigma_i$ and $\mathbf{x}=(x,\omega)$ where $\sigma_i$ is a spin function.  The one-electron integrals involving the electronic kinetic energy and the electron-nuclear attraction terms are:
\begin{equation}
	h_{pq}\equiv\int\mathrm{d}\mathbf{x} \,\chi^*_p(\mathbf{x})
		\left(-\frac{1}{2}\nabla^2-\sum_\alpha \frac{Z_\alpha}{r_{\alpha, \mathbf{x}}}\right)
					 \chi_q(\mathbf{x})
					 \label{eq:1eint}
\end{equation}
and the two-electron integrals involving the electron-electron interaction, $1/r_{12}$ are:
\begin{equation}
	h_{pqrs}\equiv  \int \mathrm{d}\mathbf{x}_{1} \mathrm{d}\mathbf{x}_{2} 
	\frac{ \chi^*_p(\mathbf{x}_{1})\chi_q^*(\mathbf{x}_{2})\chi_r(\mathbf{x}_{2})\chi_s(\mathbf{x}_{1})}{r_{1,2}}
\label{eq:2eint}
\end{equation}
In Eq.~\eqref{eq:1eint}, $\nabla^2$ is the Laplacian with respect to the electronic spatial coordinates. The positive valued scalars $r_{\alpha,\mathbf{x}}$ and $r_{1,2}$ are the Euclidean distance between the $\alpha^\textrm{th}$ nucleus and the electron and the Euclidean distance between the two electrons.  
{In both Eqs.~\eqref{eq:1eint} and~\eqref{eq:2eint}, the spin of the electron can be integrated out resulting }
{in integrals over the spatial components however since the focus is on spin orbitals,}
{the electron spin will be included in the definition of the integrals.}

\subsection{Representing the molecular Hamiltonian in terms of quantum bits}~\label{ssec:qcomp}

Just as classical computation is based on the notion of a bit, the basic unit of quantum information is the quantum bit (qubit). Qubits, being quantum, are described by a wave function instead of 
a probability distribution as in the case of classical bits. The use of the wave function 
description allows for the superposition of states and for entanglement
between different qubits. Moreover, since qubits themselves are described using 
wave functions, quantum states can be efficiently stored.

In principle, any two-level quantum mechanical system can be considered a qubit. Examples include photons, ions, and super conducting loops. Practical requirements for qubits and their manipulation was originally outlined by DiVincenzo~\cite{DiV00} and experimental progress towards satisfying the DiVincenzo criteria {were} recently reviewed~\cite{LJL+10}.  Since two-level systems can be described as spin-half particles, the relevant (Pauli) matrices are:
\begin{align*}
	&\sigma^x=\left[ \begin{array}{rr}
		0&1\\1&0
	\end{array}\right]& &\sigma^y=\left[ \begin{array}{rr}
		0&-i\\i&0
	\end{array}\right]& &\sigma^z=\left[ \begin{array}{rr}
		1&0\\0&-1
	\end{array}\right].
\label{eq:pauli}
\end{align*}
{Notice that we have defined the Pauli matrices with eigenvalues $\pm$1 instead of $\pm \hbar/2$.}
Together with the identify matrix, $\mathbf{1}$, the Pauli matrices form an operator basis for two-level systems.
{The eigenvectors of $\sigma^z$ are labeled as $|0\rangle$ and $|1\rangle$ corresponding to the eigenvalues}
{+1 and -1 respectively.}
There are several computationally equivalent models of describing quantum computation but here we focus on the circuit model of quantum computation.  A more comprehensive introduction to quantum computation can be found in Ref.~\cite{NC01}.

The quantum circuit model uses a set of elementary gates to reproduce the action of arbitrary unitary transforms.  If the elementary gates are capable of reproducing any desired unitary transform to 
 arbitrary precision, the set of gates is called universal. A universal gate set requires single qubit gates and any two-qubit entangling gate.  The two-qubit {\sf CNOT} gate leaves one qubit space {unchanged} and acts with $\sigma^x$ on the second qubit when the first qubit is in the state $\ket{1}$.  The gate is given as {\sf CNOT}$=\ket{1}\bra{1}\otimes \sigma^x + \ket{0}\bra{0}\otimes \I$.  The set, $R_x$, $R_y$, and $R_z$ gates can generate any single qubit gate {where} $R_n$ is defined as $\exp[-i \sigma^n \theta/2]$ for real $\theta$. 
			
For experimental addressability, the qubits must, in general, be distinguishable.  However, the electrons of the molecular system are indistinguishable. The Jordan-Wigner transform is used to circumvent this issue by expressing Fermionic operators in terms of the Pauli spin operators $\{\X,\Y, \Z,\I\}$ that correspond to the algebra of distinguishable spin 1/2 particles~\cite{JW28,OGKL01}. The Jordan-Wigner transform is given by:
\begin{subequations}\label{:JW}
\begin{eqnarray}
	a_j\Leftrightarrow{\I}^{\otimes j-1}\otimes \sigma^{+} \otimes {\sigma^{z}}^{\otimes N-j-1}=\left[ \begin{array}{ll}
		1&0\\0&1
	\end{array}\right]^{\otimes j-1}\otimes \left[ \begin{array}{ll}
		0&1\\0&0
	\end{array}\right]\otimes \left[ \begin{array}{rr}
		\phantom{+}1&0\\0&-1
	\end{array}\right]^{\otimes N-j-1} \label{subeq:JW(dest)}\\
    a_j^\dagger\Leftrightarrow{\I}^{\otimes j-1}\otimes \sigma^{-} \otimes {\sigma^z}^{\otimes N-j-1}=\left[ \begin{array}{ll}
		1&0\\0&1
	\end{array}\right]^{\otimes j-1}\otimes \left[ \begin{array}{ll}
		0&0\\1&0
	\end{array}\right]\otimes \left[ \begin{array}{rr}
		\phantom{+}1&0\\0&-1
	\end{array}\right]^{\otimes N-j-1}\label{subeq:JW(crea)}
\end{eqnarray}
\end{subequations}
where $\sigma^+\equiv \frac{\sigma^x+i\sigma^y}{2}=\ket{0}\bra{1}$ and $\sigma^-\equiv\frac{\sigma^x-i\sigma^y}{2}=\ket{1}\bra{0}$. The qubit state $\ket{0\dots0}$ corresponds to the vacuum state and the string of $\sigma^z$ operators, preserve the commutation relations in Eq.~\eqref{:CCR} since $\sigma^z$ and $\sigma^\pm$ anti-commute.  The spin variable representation of relevant operators after the Jordan-Wigner transformation is given in Table \ref{tbl:spin_vars}.

\section{Efficient approximations of the unitary propagator by a Trotter decomposition}\label{sec:UP}

As mentioned in the introduction, the idea of using a quantum computer for obtaining  
molecular energies requires efficiently approximating the unitary propagator, 
$\exp(-iHt)$, for a time sufficiently long to resolve the Fourier frequency to a desired 
precision.  The present section continues by describing the Trotter decomposition 
for non-commuting Hamiltonian terms and then presents the quantum circuits for 
the relevant exponentials.

\subsection{Trotter decomposition}\label{sec:Trot}

Using the second-quantized representation allows for a straightforward decomposition of the exponential map of each term of the Hamiltonian.  
However, the terms of this decomposition do not always commute.  The goal of the Trotter decomposition is to approximate the time evolution 
operator of a set of non-commuting operators.  The operators are exponentiated 
individually for small time steps and the procedure is repeated such that their product provides a reasonable approximation to the exponentiation of the sum.
Using this approximation, the construction of the time propagator can be efficiently carried out on a quantum
computer provided that the Hamiltonian can be decomposed into a sum of local Hamiltonians~\cite{Llo96}.
The first-order Trotter decomposition is given by:
\begin{equation}
e^{-iHt}= \left(e^{-ih_1\Delta t}e^{-ih_2 \Delta t}\cdots e^{-ih_N\Delta t}\right)^{\frac{t}{\Delta t}}+O(t\Delta t), 
    \label{eq:Trotter1stOrder}
\end{equation}
where $t/\Delta t$ is the Trotter number~\cite{HS05}.  As the Trotter number tends to infinity,
or equivalently $\Delta t\rightarrow 0$, the error in the approximation vanishes.
If $t/\Delta t$ is not an integer, the remainder is simulated as another Trotter time slice.
There exist higher order approximates (Suzuki-Trotter formulas) which reduce the error of approximation even further. 
For instance, the second order approximations is given by:
\begin{equation}
    e^{-iHt}\approx \left(\left(e^{-ih_1\frac{\Delta t}{2}}\cdots e^{-ih_{N-1}\frac{\Delta t}{2}}\right)e^{-ih_N\Delta t}\left(e^{-ih_{N-1}\frac{\Delta t}{2}}\cdots e^{-ih_1\frac{\Delta t}{2}}\right)\right)^\frac{t}{\Delta t}+O(t(\Delta t)^2).
    \label{eq:Trotter2ndOrder}
\end{equation}

Higher order approximations take increasingly more complicated forms \cite{HS05} and were first studied in the context of black box quantum simulation of sparse Hamiltonians by Berry \emph{et al.}~\cite{BACS06}.  They considered Hamiltonians composed of $m$ efficiently implementable terms and showed that the number of exponentials cannot scale better than linear in the time desired and the maximum frequency of the full Hamiltonian.  The proof shows that if sublinear simulation of arbitrary Hamiltonians were possible, bounds for the power of quantum computation proven in Ref.~\cite{Beals1998} could be violated leading to a contradiction.

\subsection{Quantum circuit primitives}\label{sec:qcirUP}
Each term of Eq.~\eqref{eq:ham2nd} can be exponentiated using the universal gate set described in Subsection~\ref{ssec:qcomp} after performing the Jordan-Wigner transformation to Pauli spin matrices.  We will outline the procedure for generating quantum circuits for chemical systems and summarise the results in {Table}~\ref{tblf:cir_rep}.  The construction of quantum circuits for general Fermionic Hamiltonians is further discussed in Ref.~\cite{OGKL01,SOGK+02,OHJ07}.

To understand the exponential map of the product of Pauli spin matrices, first consider the exponential map of two $\sigma^z$ operators.  To create the unitary gate $\exp[-i\frac{\theta}{2}(\Z\otimes\Z)]$, the {\sf CNOT} gate can be used to first entangle two qubits, then the $R_z$ gate is applied, {and} followed by a second {\sf CNOT} gate~\cite{NC01}.
\[\Qcircuit @C=.5cm @R=0em @!R{
&\ctrl{1} 	&\qw 		&\ctrl{1}&\qw\\
&\targ		&\gate{R_z} &\targ&\qw}
\]
This construction can be generalized to more qubits by using additional {\sf CNOT} gates.
For example, the circuit for the three-body operator involving three qubits, $\exp[{-i\frac{\theta}{2}(\sigma^{z}\otimes\sigma^{z}\otimes\sigma^{z})}]$, is simulated by the following quantum circuit:
\[\Qcircuit @C=.5em @R=0em @!R {
& \qw &    \ctrl{1}& \qw  & \qw                           &\qw     &\ctrl{1}&\qw         \\
&\qw&   \targ   & \ctrl{1}& \qw                       	 &\ctrl{1}&\targ   &\qw             \\
&\qw & \qw &      \targ   & \gate{R_z} &\targ   &\qw     &\qw                    }
\]
As seen from the three-qubit example above, this construction can be readily extended for $n$-fold products of $\sigma^z$ operators.

\subsubsection{Construction of different Pauli matrix tensor products}\label{sssec:con}
If one requires a different tensor product of Pauli matrices besides the product of $\Z$ as described in Section~\ref{sec:qcirUP}, a change of basis can be accomplished using the appropriate unitary transformation: Hadamard transformation (denoted {\sf H})  changes between $\sigma^x$ basis and $\sigma^z$ basis, and ${\sf Y}=R_x(-\pi/2)=\exp(i\sigma^x\pi/4)$ transforms from $\sigma^z$ basis to $\sigma^y$ basis and $\sf Y^\dagger$ from $\sigma^y$ to $\sigma^z$.  In matrix form,
\begin{align*}
&{\sf H}=\frac{1}{\sqrt{2}} \left[ \begin{array}{rr}
	1&1\\\phantom{-}1&-1
\end{array}\right]& &
{\sf Y}= \frac{1}{\sqrt{2}} \left[ \begin{array}{rr}
	\phantom{-}1&\phantom{-}i\\i&1
\end{array}\right].
\end{align*}
Circuits of this form are the basis for the construction of the molecular unitary propagator as illustrated in {Table}~\ref{tblf:cir_rep} where the circuit representations are given. 

\section{{The phase estimation algorithm}}\label{sec:PEA}

In this section, we describe how to obtain molecular energies given the time evolution of the molecular Hamiltonian described above.  The key idea is to Fourier transform the oscillating phase, $\bra{\psi(0)}\psi(t)\rangle=\exp(-iEt)$, to obtain the electronic energy.  The eigenenergy is converted into relative phases. The relative phase can then be measured using the quantum phase
estimation algorithm (PEA). As the phase is measured the input state partially collapses to the set of states consistent with the measurements obtained up to that point.

To determine an eigenvalue associated with an eigenstate, consider the phase of an eigenstate of the Hamiltonian, $H$, evolving dependent on a register qubit: 
\begin{equation}
	|0\rangle\ket{\psi_n}+e^{-iHt}|1\rangle\ket{\psi_n}=|0\rangle\ket{\psi_n}+e^{-iE_{n}t}\ket{1}\ket{\psi_n}.
	\label{eqn:SE}
\end{equation}
By letting $E_{n}=2\pi(\phi-K)/t$ where $0\leq\phi<1$ and $K$ is an unobservable integer,
the unknown eigenvalue becomes encoded in the relative phase of the register qubit quantum
state as $\ket{0}+e^{-2\pi i(\phi-K)}\ket{1}$~\cite{PP00,AGDL+05,Kit95}.  Then, $\phi$ is estimated using the phase estimation technique detailed in Subsection~\ref{ssec:ipea}.

Once $\phi$ has been estimated, it must be inverted to obtain the energy. Given bounds for the energy eigenvalue, $[E_{min},E_{max})$, the time of evolution is selected as $t=2\pi/(E_{max}-E_{min})=2\pi/\omega$  and an energy shift of $E_s$ is used to ensure $K=(E_s-E_{min})/\omega$ is an integer. 
The energy shift, $E_s$, is effected by a gate on the register qubit which applies a phase to the qubit if it is in state $\ket{1}$ and does nothing otherwise. Using these parameters the measured value of $\phi$ corresponds to the value of the energy, $E_\phi=\omega(\phi -K)+E_s$. 

The PEA can be accomplished in several different ways depending on the technology available and in the recent experimental implementations, the iterative phase estimation algorithm~\cite{PP00,TN06,DJSW06,XMJ+07} was used.  We present it here for completeness.

\subsection{{Iterative Phase Estimation}}\label{ssec:ipea}

The goal of the phase estimation algorithm is to obtain the best $L$ bit estimate of a phase $\phi$.  The readout will be values of 0 or 1 from the fractional binary expansion of $\phi$:
\begin{equation}
	\phi=0.j_0j_1j_2\dots j_{L-1}=\left(\frac{j_0}{2}\right)+\left(\frac{j_1}{4}\right)+\left(\frac{j_2}{8}\right)+\dots+\left(\frac{j_{L-1}}{2^{L}}\right).
	\label{:Bin}
\end{equation}
Each $j_k$ is zero or one.  The relative phase generated by the unitary propagator $U(t)=\exp(-iHt)$ applied as in \eqref{eqn:SE} is defined as $\exp(2\pi i \phi)$.  Performing $U(2^kt)$ results in a relative phase $\exp(2\pi i (2^k\phi))$.  Observing that $2^k\phi=j_0j_1\dots j_{k-1}.j_kj_{k+1}\dots$ we can finite binary expansion of length $L$ deterministically.

First, $U(2^Lt)$ is implemented resulting in relative phase $\exp(2\pi i \; j_0\dots j_{L-1}.j_L)=\exp(2\pi i\; j_L/2)$.  Since $j_L$ is 0 or 1, the relative phase is either +1 or -1, respectively.  The Hadamard transform described in Section~\ref{sssec:con} distinguishes $\ket{0}+\ket{1}$ from $\ket{0}-\ket{1}$ allowing a projective measurement to identify $j_L$.

To obtain $j_{L-1}$ and more significant bits, the iterative method uses the gate $\sf S_k$ to deterministically obtaining $j_k$.  Since data is read from the least significant digit first, the counter-rotation $\sf S_k$ is a function of previously obtained bits. The form of the $\sf S_k$ gate is:
\[
{\sf S_k}=
\left[\begin{tabular}{cc}
	1&0\\
	0&$\Phi_k$
\end{tabular}\right],
\quad\mathrm{with}\quad
\Phi_k=\exp\left[2\pi i\sum_{l=2}^{L-k+1}\frac{j_{k+l-1}}{2^{l}}\right].
\]
This gate removes the $L-k-1$ least significant digits so the state of quantum computer becomes $(\ket{0}+\exp[-i\pi j_k]\ket{1})\ket{\psi_n}$ where $j_k$ is zero or one.  Finally, effecting the Hadamard transformation leads to $\ket{j_k}\ket{\psi_n}$ and measurement of the register in the $\{\ket{0},\ket{1}\}$ basis yields the value of $j_k$.  When the binary expansion of $\phi$ is length $L$, the measurements are deterministic, otherwise the remainder causes some randomness but does not significantly affect the results.  A complete error analysis can be found in Refs.~\cite{NC01,DJSW06}.


As noted in Ref.~\cite{BCC06}, in phase estimation algorithm the number of uses of $U(t_0)$ for a fixed time $t_0$ scales exponentially with the number of bits desired from $\phi$. This is a consequence of the Fourier uncertainty principal; the more information required in the frequency domain the longer the propagation time.  When the unitary is decomposed into gates, this means an exponential increase in the gates is required for an exponential increase in the precision of the measurement.    
\begin{figure}[t]
\[
\centerline{\Qcircuit @C=1em @R=1em @!R {
	\lstick{\ket{0}}	              	&\gate{\sf H}	&\ctrl{1}      	&\gate{\sf S_k} &\gate{\sf H}	& \meter 	 &\rstick{j_k} \cw\\
   	\lstick{\ket{\psi}}	&\qw      	&\gate{U^{2^k}}	&\qw      	&\qw		&\qw 		 &\rstick{\ket{\psi}} \qw
        }}
	\]
	\caption{
	 Iterative phase estimation circuit for obtaining the $k^{th}$ bit. The phase is represented using a binary decimal expansion $\phi=0.j_0j_1j_2j_3\cdots j_n$.  To gain precision in the algorithm, one adjusts the parameterized gate ${\sf S_k}$ according to the values of all previous obtained bits. This $U$ is a representation of the propagator of the system being simulated.  Acting on the eigenstate with this operator advances the phase of the state allowing bit $j_k$ to be read out deterministically.
	}
	\label{fig2}
\end{figure}
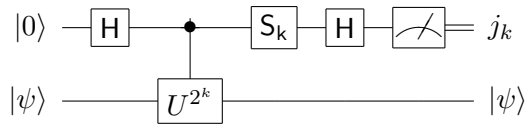

To produce the state indicated in \eqref{eqn:SE}, the evolution of the eigenstate must be dependent on the register qubit requiring that the construction of the unitary evolution operator described in Section~\ref{sec:qcirUP} be modified. The constructions listed in {Table}~\ref{tblf:cir_rep} only need to be slightly modified; since the underlying constructions rely on $R_z$ gates, changing these rotations into controlled $R_z$ rotations ($|1\rangle\langle 1|\otimes R_z+|0\rangle\langle 0|\otimes \mathbf{1}$) is sufficient to make the entire unitary dependent on the readout qubit.


To obtain ground state energies, PEA relies on the assumption that the input wave function has signification overlap with the ground state. Since each qubits represents the occupancy of molecular orbitals in the $N$-electron wave function, the HF guess for the ground state $\ket{\psi_0^{HF}}$ requires no superposition of states and is thus straightforward to prepare. For $|\bra{\psi_0^{HF}}\psi_0^{FCI}\rangle|=1-\epsilon$, where $\epsilon$ is small, the phase estimation algorithm can be applied to retrieve an estimate of the ground state energy.  Simultaneously, the state of the system will collapse to $\ket{\psi_0^{FCI}}$ when measured in the $H^{FCI}$ basis (via PEA) with high probability~\cite{PP00,Kit95}. If the Hartree-Fock guess is insufficient, more sophisticated state preparation procedures exist and these were reviewed in Ref.~\cite{Kassal2010}.  The adiabatic scheme for state preparation~\cite{AGDL+05} is analyzed in the following section.


\section{Adiabatic state preparation}\label{appx:StPrep}
When the initial state prepared as input has low overlap with the ground
state, the initial state must be improved so that PEA will collapse the input to
the correct state with greater efficiency.
This section explains the method of preparing an input state to the simulation algorithm using adiabatic quantum 
computation~\cite{FGGS00,WBL02,AGDL+05}. 
First, consider the Hartree-Fock wave function as an approximation to the FCI wave function.  This is the output of an classical algorithm returning, in polynomial time, a computational basis state where the qubits which correspond to occupied molecular orbitals are in the state $\ket{1}$ with the remaining qubits in state $\ket{0}$. 
To increase overlap of the wave function, after the system is prepared in state $\ket{\psi_0^{HF}}$, the Hamiltonian $ {H}_{FCI}$ is slowly applied and the actual ground state is recovered by adiabatic evolution.  Consider a smooth one-parameter family of adiabatic path Hamiltonians,
\begin{equation}
{H(s)}=(1-s) {H}_{HF}+s {H}_{FCI},
\label{eq:adiabatic_path}
\end{equation}
for monotonic $s\in[0,1]$.  This was the adiabatic path originally proposed by us in Ref.~\cite{AGDL+05}.  Other paths may be used as in Ref.~\cite{PVAA+08} but in this study we restrict our attention to evolution of the form in Eq.~(\ref{eq:adiabatic_path}).

Let the instantaneous energies of $H(s)$ be given by the sequence,
\begin{equation}
E_0(s)< E_1(s)\leq\cdots\leq E_{N-1}(s),
\end{equation}
then the adiabatic state preparation procedure is efficient whenever the total run
time, $T$, satisfies the following:
\begin{equation}
T\gg\min_{0\leq s\leq 1} \left(E_1(s)-E_0(s)\right)^{-2},
\end{equation}
according to known results relating the adiabatic theorem to complexity theory~\cite{FGGS00}.  After adiabatic evolution, the state of the system is $\ket{\psi_0^{FCI}}$, which is the ground state of the molecular Hamiltonian $ {H}_0^{FCI}$.  Modified versions of this procedure exist using decoherence to achieve faster evolution and are discussed in Refs.~\cite{BKS09,WA08}.

Assume that the adiabatic evolution induced transitions into higher energy states and so the un-normalized state of the system is $\ket{\psi_0^{FCI}}+\lambda\ket{k}$, where $\ket{k}\in\mathcal{H}= 1-\ket{\psi_0^{FCI}}\bra{\psi_0^{FCI}}$. While the error in the wave function is linear in $\lambda<1$, the overestimate of the energy in the expectation value $\langle {H}_0^{FCI}\rangle$ is only quadratic.

It is unclear how this method will scale. It is possible to prepare a desired
ground state efficiently provided that the gap between the ground and excited states is
sufficiently large~\cite{FGGS00}. This depends on the adiabatic path taken.  Finding the ground state energy of a random Hamiltonian, even for simple models, 
is known to be complete for the quantum analogue of the class NP~\cite{KKR06,OT06}.  

There are other ways to perform state preparation, for example, by going beyond Hartree-Fock theory~\cite{WKAG+08,WAF09}. A broader discussion of state preparation for quantum simulation can be found in Refs.~\cite{Kassal2010,WKA09}.  Recently, in Ref.~\cite{Veis2010}, the effects of initial states for the phase estimated quantum simulation CH$_2$ molecules {were} studied for a variety of geometries and eigenstates using initial guesses obtained via multi-configuration approaches.

\section{Simulating the hydrogen molecule}\label{sec:H2}

To illustrate the algorithmic details of a scalable simulation of quantum systems, the hydrogen molecule in a minimal basis is used as an instructive example. The minimal basis is the minimum number of spatial-functions needed to describe the system and in the case of H$_2$, one spatial-function is needed per atom denoted $\varphi_{H1}$ and $\varphi_{H2}$.  In this simple case, the Hartree-Fock procedure is not necessary as the molecular spatial-orbitals are determined by symmetry and are given by $\varphi_{g}=\varphi_{H1}+\varphi_{H2}$ and $\varphi_{u}=\varphi_{H1}-\varphi_{H2}$. These two spatial functions correspond to four orbitals that will be identified as:
\begin{align}
\ket{\chi_1}&=\ket{\varphi_g}\ket{\alpha},&\ket{\chi_2}&=\ket{\varphi_g}\ket{\beta},&\ket{\chi_3}&=\ket{\varphi_u}\ket{\alpha},&
\ket{\chi_4}&=\ket{\varphi_u}\ket{\beta}.
	\label{eq:spinorbs}
\end{align}

The form of the spatial function is determined by the basis set used.  The STO-3G basis is a common minimal basis that approximates a single electron spatial Slater type orbitals (STO), with a contraction of three real Gaussian functions~\cite{SO96}.
Using this orbital basis, the spatial integrals of the Hamiltonian were evaluated in Table~\ref{tbl:Int} for H$_2$ at bond distance 1.401000 atomic units (7.414$\cdot 10^{-11}$ m).

\begin{table}[t]
	\tbl{The one-electron and two-electron integrals defined in Eqs.~(\ref{eq:1eint}) and (\ref{eq:2eint}) are evaluated using the molecular spatial orbitals obtained from a restricted Hartree-Fock calculation at an internuclear distance of 1.401000 atomic units (7.414$\cdot 10^{-11}$ m)~\cite{PyQ}.}	
{\begin{tabular}{c @{\hspace{1cm} }c}
		\toprule
		Integrals & Value (a.u.)\\
		\colrule
		$h_{11}=h_{22}$&-1.252477\\
		$h_{33}=h_{44}$&-0.475934\\[3ex]
		$h_{1221}=h_{2112}$& 0.674493\\[3ex]
		$h_{3443}=h_{4334}$& 0.697397\\[3ex]
		\begin{tabular}{r}
		$\phantom{=}h_{1331}=h_{1441}=h_{2332}=h_{4224}$\\
		$=h_{3113}=h_{3223}=h_{4114}=h_{4224}$
		\end{tabular}	& 0.663472\\[3ex]
		$h_{1313}=h_{2424}= h_{3241}=h_{1423}=h_{1243}$& 0.181287\\
		\botrule
	\end{tabular}}
	\label{tbl:Int}
\end{table}

Considering $H$ from Eq.~\eqref{eq:ham2nd} as $H^{(1)}+H^{(2)}$ we have{,}
\begin{equation}
	H^{(1)}=h_{11}\ad_1a_1+h_{22}\ad_2a_2
		+h_{33}\ad_3a_3+h_{44}\ad_4a_4
	\label{eq:H2singles}
\end{equation}
The following circuit applies the single-electron propagator for
a time $t$:
\[
\tiny\Qcircuit @C=.25em @R=.5em @!R {
\lstick{PEA}& \qw 	& \ctrl{1}	   	&\ctrl{2}	& \ctrl{3} 	& \ctrl{4}&\qw\\
\lstick{\chi_1}	& \qw 	& \gate{ {\sf T}(h_{11}t)} 	&\qw     	& \qw     	&\qw&\qw\\
\lstick{\chi_2}& \qw 	& \qw     		&\gate{{\sf T}(h_{22}t)}& \qw    	&\qw	&\qw\\
\lstick{\chi_3}& \qw 	& \qw     		& \qw   	& \gate{{\sf T}(h_{33}t)}&\qw	   &\qw\\
\lstick{\chi_4}& \qw 	& \qw     		& \qw     	& \qw  		 & \gate{{\sf T}(h_{33}t)}&\qw
}
\]
The gate $\sf T$ is defined as:
\begin{equation}
	\boxed{\sf T(\theta)}=\left[ \begin{array}{cc}
		1&\\
		&e^{-i\theta}
	\end{array}\right].
	\label{eq:numops}
\end{equation}

The two electron Hamiltonian can also be expanded. As electrons are indistinguishable,  
\[h_{pqrs}=\int \textrm{d}\bfx_1\:\textrm{d}\bfx_2 \frac{\chi_p(\bfx_1)\chi_q(\bfx_2)\chi_r(\bfx_2)\chi_s(\bfx_1)  }{r_{12}}
	  =\int \textrm{d}\bfx_2\:\textrm{d}\bfx_1 \frac{\chi_p(\bfx_2)\chi_q(\bfx_1)\chi_r(\bfx_1)\chi_s(\bfx_2)  }{r_{12}}
	  =h_{qpsr},\]
and $\ad_p\ad_qa_ra_s=\ad_q\ad_pa_sa_r$, the two electron Hamiltonian can be simplified as:
\begin{eqnarray}
	H^{(2)}&=&h_{1221}\ad_1\ad_2a_2a_1+
	h_{3443}\TwoE{3}{4}{4}{3}+h_{1441}\TwoE{1}{4}{4}{1}	+h_{2332}\TwoE{2}{3}{3}{2}\nonumber\\
	&&+\left(h_{1331}-h_{1313}\right)\TwoE{1}{3}{3}{1} +\left( h_{2442} -h_{2424}\right)\TwoE{2}{4}{4}{2}\nonumber\\
	&&+
	 \Re(h_{1423})(\TwoE{1}{4}{2}{3}+\TwoE{3}{2}{4}{1})+\Re(h_{1243})(\TwoE{1}{2}{4}{3}+\TwoE{3}{4}{2}{1})\nonumber\\
	&& +\Im(h_{1423})(\TwoE{1}{4}{2}{3}-\TwoE{3}{2}{4}{1})+\Im(h_{1243})(\TwoE{1}{2}{4}{3}-\TwoE{3}{4}{2}{1})
	\label{eq:H2e}
\end{eqnarray}
The first six terms,  
\begin{eqnarray*}
	 &&h_{1221}\ad_1\ad_2a_2a_1+h_{3443}\TwoE{3}{4}{4}{3}+h_{1441}\TwoE{1}{4}{4}{1}\\
	&+&h_{2332}\TwoE{2}{3}{3}{2}+\left(h_{1331}-h_{1313}\right)\TwoE{1}{3}{3}{1} +\left( h_{2442} -h_{2424}\right)\TwoE{2}{4}{4}{2}
\end{eqnarray*}
can be simulated using the system Hamiltonian that employs only commuting two-local terms described in Section~\ref{sec:UP}.
Notice after the Jordan-Wigner transform of the relevant operator we have:
\begin{eqnarray}
\sum_{p<q}(h_{pqqp}-h_{pqpq}\delta_{\sigma_p\sigma_q})\ad_p\ad_qa_q a_p
			   &=& \left(\frac{1}{4}\sum_{p<q}(h_{pqqp}-h_{pqpq}\delta_{\sigma_p\sigma_q})\right)\I\nonumber\\
			     && -\frac{1}{4}\sum_{q}\left(\sum_{p\neq q}(h_{pqqp}-h_{pqpq}\delta_{\sigma_p\sigma_q})\right)\Z_q\nonumber\\
			     &&+\sum_{p<q}\frac{(h_{pqqp}-h_{pqpq}\delta_{\sigma_p\sigma_q})}{4}\Z_p\Z_q.
	\label{eq:2eSimple}
\end{eqnarray}
The factor of $1/2$ is accounted for because the indistinguishably of the electrons reduces the summation in Eq.~\eqref{eq:ham2nd}.

Following Eq.~\eqref{eq:2eSimple}, let $\Theta\equiv(1/4)\sum_{p<q}(h_{pqqp}-h_{pqpq}\delta_{\sigma_p\sigma_q})$ and $\theta_p\equiv \sum_{q:p\neq q}(h_{pqqp}-h_{pqpq}\delta_{\sigma_p\sigma_q}).$ Then following circuit illustrates the one and two-local interactions required to implement Eq.~\eqref{eq:2eSimple}:
\[
\tiny\Qcircuit @C=.25em @R=.5em @!R {
\lstick{PEA}& \gate{ {\sf T}(\Theta t)}& \ctrl{1}	   	&\ctrl{2}		& \ctrl{3} 		& \ctrl{4}		&\qw\\
\lstick{\chi_1}	&\qw 		& \gate{R_z(\theta_1t)} &\qw     		& \qw     		&\qw			&\qw\\
\lstick{\chi_2}& \qw 		& \qw     		&\gate{R_z(\theta_2t)}	& \qw    		&\qw			&\qw\\
\lstick{\chi_3}& \qw 		& \qw     		& \qw   		& \gate{R_z(\theta_3t)}	&\qw	   		&\qw\\
\lstick{\chi_4}& \qw 		& \qw     		& \qw     		& \qw  		 	& \gate{R_z(\theta_4t)}	&\qw
} 
\]
Defining $\eta_{pq}\equiv\frac{1}{4}(h_{pqqp}-h_{pqpq}\delta_{\sigma_p\sigma_q})$, the three local interactions can be depicted as:
\begin{equation*}
\tiny\Qcircuit @C=.25em @R=.5em @!R {
\lstick{PEA}   &\qw						&\ctrl{4}	&\qw		&\qw					&\ctrl{4}	&\qw						&\qw						&\ctrl{4}	&\qw		&\qw						&\ctrl{3}	&\qw		&\qw		&\ctrl{3}	&\qw		&\qw		&\ctrl{2}		&\qw		\\
\lstick{\chi_1}&\qw						&\qw		&\qw		&\qw					&\qw		&\qw						&\ctrl{3}                                  &\qw		&\ctrl{3}	&\qw						&\qw		&\qw   		&\ctrl{2}       &\qw       	&\ctrl{2}	&\ctrl{1}                                  	&\qw			&\ctrl{1}	\\
\lstick{\chi_2}&\qw						&\qw		&\qw		&\ctrl{2}                                &\qw		&\ctrl{2}                                  	&\qw						&\qw		&\qw		&\ctrl{1}                                  	&\qw		&\ctrl{1}	&\qw		&\qw		&\qw		&\targ		&\gate{\scriptstyle R_z(\eta_{12}t)}	&\targ		\\
\lstick{\chi_3}&\ctrl{1}		&\qw		&\ctrl{1}	&\qw					&\qw		&\qw						&\qw						&\qw		&\qw		&\targ						&\gate{\scriptstyle R_z(\eta_{23}t) }	&\targ		&\targ		&\gate{\scriptstyle R_z(\eta_{13}t) }	&\targ		&\qw		&\qw			&\qw		\\
\lstick{\chi_4}&\targ						&\gate{\scriptstyle  R_z(\eta_{34}t)	}	&\targ		&\targ					&\gate{\scriptstyle R_z(\eta_{24}t) }	&\targ						&\targ						&\gate{\scriptstyle R_z(\eta_{14}t) }	&\targ		&\qw						&\qw		&\qw		&\qw		&\qw		&\qw		&\qw		&\qw			&\qw
}
\end{equation*}
Each term commutes thus can be realized in any order.

The remaining terms are strictly real leaving:
\[h_{1423}(\TwoE{1}{4}{2}{3}+\TwoE{3}{2}{4}{1})+(h_{1243})(\TwoE{1}{2}{4}{3}+\TwoE{3}{4}{2}{1}).\]
Since the orbitals are real, the integrals  $h_{1243}=h_{1423}$  are equivalent. Therefore, we are left with the task of simulating $2h_{1423}(\TwoE{1}{4}{2}{3}+\TwoE{3}{2}{4}{1})$. 

Consider the general term $\ad_p\ad_qa_r a_s+\ad_s\ad_ra_q a_p$.  Due to the anti-commutation rules, all sets of operators
corresponding to a set of four distinct spin-orbitals, ($p$, $q$, $r$, $s$), are simulated using the same circuit.
This is due to the fact that the Jordan-Wigner of each operator generates same set of operators (namely, the eight 
combinations involving an even number of $\X$ and $\Y$ operators).  However, depending on if $\sigma^+$ or $\sigma^-$ is 
used each term of spin operators will have a different sign. If we define:
\begin{eqnarray}
	h^{(1)}&=& (h_{pqrs}\delta_{\sigma_p\sigma_s}\delta_{\sigma_q\sigma_r}-h_{qprs}\delta_{\sigma_p\sigma_r}\delta_{\sigma_q\sigma_s})\label{eq:h1}\\
	h^{(2)}&=& (h_{psqr}\delta_{\sigma_p\sigma_r}\delta_{\sigma_q\sigma_s}-h_{spqr}\delta_{\sigma_p\sigma_q}\delta_{\sigma_r\sigma_s})\label{eq:h2}\\
	h^{(3)}&=& (h_{prsq}\delta_{\sigma_p\sigma_q}\delta_{\sigma_r\sigma_s}-h_{prqs}\delta_{\sigma_p\sigma_s}\delta_{\sigma_q\sigma_r})\label{eq:h3}
\end{eqnarray}
then
\begin{eqnarray}
	\frac{1}{2}\sum {h}_{pqrs}\ad_p\ad_qa_r a_s+\ad_s\ad_ra_q a_p&=&\frac{1}{8}\left(\bigotimes_{k=p+1}^{q-1}\bigotimes_{k=r+1}^{s-1} \Z_k\right)\\
		&&	\left(
	\begin{tabular}{l}
		$\phantom{+} (\X_p\X_q\X_r\X_s+\Y_p\Y_q\Y_r\Y_s)(-h^{(1)}-h^{(2)}+h^{(3)})$\\
		$+(\X_p\X_q\Y_r\Y_s+\Y_p\Y_q\X_r\X_s)(+h^{(1)}-h^{(2)}+h^{(3)})$\\
		$+(\Y_p\X_q\Y_r\X_s+\X_p\Y_q\X_r\Y_s)(-h^{(1)}-h^{(2)}-h^{(3)})$\\
		$+(\Y_p\X_q\X_r\Y_s+\X_p\Y_q\Y_r\X_s)(-h^{(1)}+h^{(2)}+h^{(3)})$
		 \end{tabular} \right).\nonumber
	\label{eq:EEop}
\end{eqnarray}

Applying this to the hydrogen molecule, observe that $h^{(1)}=-h^{(2)}$ and $h^{(3)}=0$ indicating that only the terms $\{\X\X\Y\Y,\Y\Y\X\X,\Y\X\X\Y,\X\Y\Y\X\}$ must be considered. The resulting quantum circuit is illustrated in Table~\ref{cir:4func}.

To assess the Trotter error, we simulated this circuit using a classical computer using the first-order Trotter decomposition. The pseudo-code for the H$_2$ simulation is given in Appendix~\ref{appx:code} and the results are summarized in Fig.~3. Although the gates increase with the Trotter number, reducing the Trotter error of the dynamics is only practical if the measurement is precise enough to detect such errors.  Thus, in practice, there is a balance between the Trotter number selected and the number of bits to be obtained by the measurement procedure. 
Finally, as noted in section~\ref{sec:Trot}, higher-order Trotter decompositions also allow 
for more precise simulation with larger time steps.  In fact, given the largest 
eigenvalue of the simulated system as $\lambda$ the number of gates for simulation
time $t$  is $O(t \lambda)$ using the appropriate order of Trotter decomposition~\cite{BACS06}. 

\begin{figure}
	\begin{center}
\begin{minipage}{110mm}
	\subfigure[ Plot of gates to simulate the H$_2$ versus time step used in the first order Trotter approximation. ]{\resizebox*{5cm}{!}{\includegraphics[width=5.3cm]{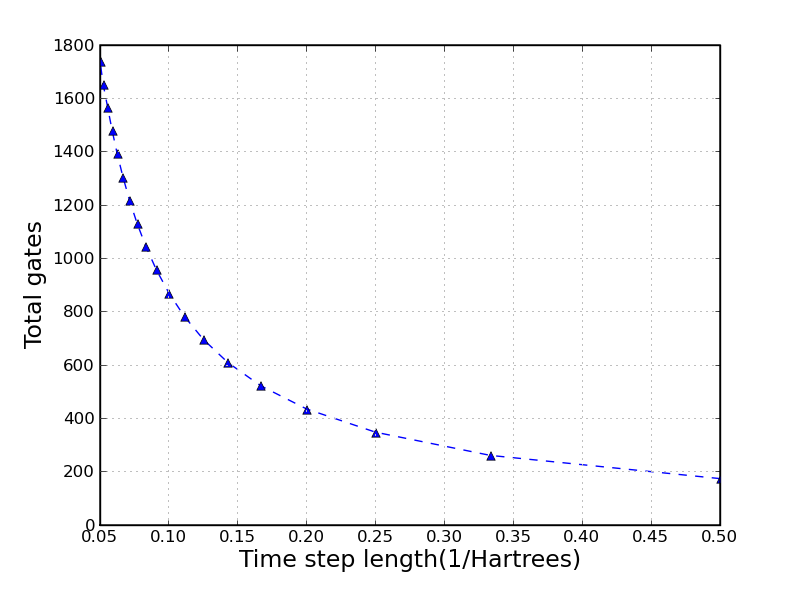}}}
	\subfigure[ Plot of relative error of approximation as a function of gates used.  ]{\resizebox*{5cm}{!}{\includegraphics[width=5cm]{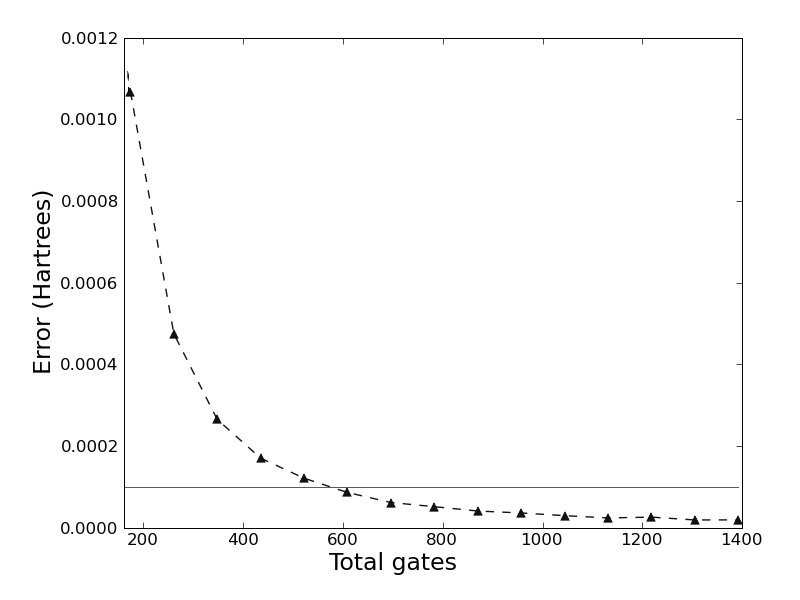}}}
	\caption{ The unitary propagator corresponding to this Hamiltonian is approximated using a first order Trotter decomposition and these graphs provide analysis of the Trotter error and the number of gates used at each Trotter number, $T_n$. The unitary propagator is simulated by applying each small term of the second quantized Hamiltonian for small time steps $dt$ and repeating the process $T_n=t/dt$ times.  As $dt$ decreases the error in the approximation decreases at the expense of more gates. Zero error represents the eigenvalue of the Hamiltonian of H$_2$ in the minimal basis at a separation of 1.4 atomic units.       The horizontal line of \emph{(b)} represents the threshold for energy error of $10^{-4}$ atomic units.  }
	\label{fig:gates}
\end{minipage}
\end{center}
\end{figure}

\section{Conclusions}\label{sec:conclusion}

In this paper, we mapped the full configuration interaction (FCI) method from quantum chemistry into a quantum algorithm.  We reviewed the electronic structure problem, techniques of creating the simulated propagator, and explicitly illustrated this construction for H$_2$ for a single time step of a first-order Trotter expansion. 

Applicability of quantum simulation comes down to the ability to propagate the simulated system with a specified error tolerance.  Since phase estimation is essentially a Fourier transform of the frequency of phase oscillations (which are proportional to the eigenenergy) to obtain more precise determination of the frequency, a longer propagation time is necessary.  The longer simulation time requires more manipulations of the computational system. 

In the coming future, small scale experiments such as the simulation of the circuits we have presented for H$_2$ on a quantum computer will likely be possible.  Experimental realizations of quantum chemistry on quantum devices have only recently been achieved~\cite{LW+09,DXP+09}. We hope that the present paper will continue the interest by giving an example of a scalable construction of the unitary propagator for the H$_2$  molecule in an explicit form, which poses the next logical challenge for experimental realization.

\section*{Acknowledgements} {\small We thank H. Wang, A. Dutoi, P. Love, M. Mohseni, B. Lanyon and A. White.
This work received funding from the Faculty of Arts and Sciences at Harvard University, Engineering and Physical Sciences Research Council grant EP/G003017/1 (JDB), Defense Advance Research Projects Agency under the Young Faculty Award (N66001-09-1-2101) (AA-G), the Camille and
Henry Dreyfus Foundation (AA-G), the Sloan Foundation (AA-G), NSF CCI 1037992-CHE, and the Army Research Office under contract W911-NF-07-0304 (JDW, AA-G).}

\newpage

\appendices
\section{{Example program for a quantum computer to calculate electronic energies for the hydrogen molecule} }\label{appx:code}

The time evolution of input states is measured to extract eigenvalues of the system.  A state preparation procedure must be used if a particular state is desired (e.g. the ground state).  Time evolution is implemented in such a way that a measurable relative phase is generated on a register qubit.  This register qubit is then measured to obtain a bit of information about the phase.  The time evolution is then effected for half as long and again the register is measured to obtain the next digit.  This process is repeated until all digits are read out and then the phase can be converted to an energy eigenvalue following the prescription of Section~\ref{sec:PEA}.  

In this appendix, we explicitly spell out the gates that would be necessary for effecting a single time step of the evolution generated by the Hydrogen molecule's Hamiltonian.   The time step idea is to alternatively apply each of the non-commuting unitary operators for a fraction of the total time and repeat the time steps until the total time has been reached.  As the length of the time steps goes to zero the error of the approximation vanishes.  For constructing long time evolution the time steps should be composed following the Trotter prescription~\ref{sec:Trot}

The decomposition of the propagator of H$_2$ Hamiltonian is 
given using standard quantum gates. The entire unitary 
operator is applied if the register qubit is in state $|1\rangle$.
Qubits are named: Register, Q1, Q2, Q3, Q4.
The integrals $h_{ij}$ and $h_{ijkl}$ are parametrized by the nuclear configuration and  are given in Table~\ref{tbl:Int} for the for the equilibrium geometry.
Additionally, we define $n_{ij}=(h_{ijji}-h_{jiji})/4$.  Note only $h_{1313}$ and $h_{2424}$ are the only non-zero terms due to spin orthogonality.
Following the main text, let $\Theta\equiv(1/4)\sum_{p<q}(h_{pqqp}-h_{pqpq}\delta_{\sigma_p\sigma_q})$ and $\theta_p\equiv \sum_{q:p\neq q}(h_{pqqp}-h_{pqpq}\delta_{\sigma_p\sigma_q}).$  

\begin{tabular}[h]{llll}
\\Gate\phantom{jkjljlhnbbvvce}  & Target qubit\phantom{space} & Control qubit\phantom{space}& Parameter\\
\toprule
Hadamard& Register &&\\
\colrule
\multicolumn{4}{l}{\textsc{\underline{Single Electron Operators}}}\\
cPhase & Q1 & Register &  $h_{11}  t$\\ 
cPhase & Q2 & Register &  $h_{22}  t$\\ 
cPhase & Q3 & Register &  $h_{33}  t$\\ 
cPhase & Q4 & Register &  $h_{44}  t$\\ 
\colrule
\multicolumn{4}{l}{\textsc{\underline{Two Electron Operators: number-number operator}}}\\
Phase & Register &&$\Theta t$\\
cR$_z$ &Q1&Register  &-$\theta_1 t$\\
cR$_z$ &Q2&Register  &-$\theta_2 t$\\
cR$_z$ &Q3&Register  &-$\theta_3 t$\\
cR$_z$ &Q4&Register  &-$\theta_4 t$\\
cNot   &Q4 &Q3 &\\
cR$_z$ &Q4 &Register &$2n_{34}t$\\
cNot   &Q4 &Q3 &\\
cNot   &Q4 &Q2 &\\
cR$_z$ &Q4 &Register &$2n_{24}t$\\
cNot   &Q4 &Q2 &\\
cNot   &Q4 &Q1 &\\
cR$_z$ &Q4 &Register &$2n_{14}t$\\
cNot   &Q4 &Q1 &\\
cNot   &Q3 &Q2 &\\
cR$_z$ &Q3 &Register &$2n_{23}t$\\
cNot   &Q3 &Q2 &\\
cNot   &Q3 &Q1 &\\
cR$_z$ &Q3 &Register &$2n_{13}t$\\
cNot   &Q3 &Q1 &\\
cNot   &Q2 &Q1 &\\
cR$_z$ &Q2 &Register &$2n_{12}t$\\
cNot   &Q2 &Q1 \\
\colrule
\multicolumn{4}{l}{\underline{\textsc{Two Electron Operators: excitation-excitation operator}}}\\
\textsc{XXYY} & & &\\
Hadamard & Q1 & &\\
Hadamard & Q2 & &\\
R$_x$ &Q3 &&-$\pi/2$\\
R$_x$ &Q4 &&-$\pi/2$\\
cNot& Q2 &Q1&\\
cNot& Q3 &Q2&\\
cNot& Q4 &Q3&\\
cR$_z$ &Q4&Register&-$t(h_{1423}+h_{1243})/4$\\
cNot& Q4 &Q3&\\
cNot& Q3 &Q2&\\
cNot& Q2 &Q1&\\
R$_x$ &Q4 &&$\pi/2$\\
R$_x$ &Q3 &&$\pi/2$\\
Hadamard & Q2 & &\\
\botrule
\multicolumn{4}{r}{\small \sl continued on next page}
\end{tabular}
\newpage
\begin{tabular}{llll}
Gate\phantom{jkjljlhnbbvvce} & Target qubit\phantom{space} & Control qubit\phantom{space}& Parameter\\
\toprule
Hadamard & Q1 & &\\
\textsc{YYXX} & & &\\
R$_x$ &Q1 &&-$\pi/2$\\
R$_x$ &Q2 &&-$\pi/2$\\
Hadamard & Q3 & &\\
Hadamard & Q4 & &\\
cNot& Q2 &Q1&\\
cNot& Q3 &Q2&\\
cNot& Q4 &Q3&\\
cR$_z$ &Q4&Register&-$t(h_{1423}+h_{1243})/4$\\
cNot& Q4 &Q3&\\
cNot& Q3 &Q2&\\
cNot& Q2 &Q1&\\
Hadamard & Q4 & &\\
Hadamard & Q3 & &\\
R$_x$ &Q2 &&$\pi/2$\\
R$_x$ &Q1 &&$\pi/2$\\
\textsc{XYYX} & & &\\
Hadamard & Q1 & &\\
R$_x$ &Q2 &&-$\pi/2$\\
R$_x$ &Q3 &&-$\pi/2$\\
Hadamard & Q4 & &\\
cNot& Q2 &Q1&\\
cNot& Q3 &Q2&\\
cNot& Q4 &Q3&\\
cR$_z$ &Q4&Register&$t(h_{1423}+h_{1243})/4$\\
cNot& Q4 &Q3&\\
cNot& Q3 &Q2&\\
cNot& Q2 &Q1&\\
Hadamard & Q4 & &\\
Hadamard & Q3 & &\\
R$_x$ &Q2 &&$\pi/2$\\
R$_x$ &Q1 &&$\pi/2$\\
\textsc{YXXY} & & &\\
R$_x$ &Q1 &&$\pi/2$\\
Hadamard & Q2 & &\\
Hadamard & Q3 & &\\
R$_x$ &Q4 &&$\pi/2$\\
cNot& Q2 &Q1&\\
cNot& Q3 &Q2&\\
cNot& Q4 &Q3&\\
cR$_z$ &Q4&Register&$t(h_{1423}+h_{1243})/4$\\
cNot& Q4 &Q3&\\
cNot& Q3 &Q2&\\
cNot& Q2 &Q1&\\
R$_x$ &Q1 &&$-\pi/2$\\
Hadamard & Q2 & &\\
Hadamard & Q3 & &\\
R$_x$ &Q4 &&$-\pi/2$\\
\botrule
\end{tabular}

\bibliographystyle{tMPH}
\bibliography{h2gts}

\begin{thebibliography}{49}
\providecommand{\url}[1]{\texttt{#1}}
\providecommand{\urlprefix}{URL }
\markboth{Whitfield et. al}{arXiv:1001.3855}

\bibitem{SO96}
A. Szabo and N. Ostlund, \emph{Modern Quantum Chemistry: Introduction to
  Advanced Electronic Structure Theory}   (Dover Publications, Mineola, NY,
  1996).

\bibitem{HJO00}
T. Helgaker, P. Jorgensen and J. Olsen, \emph{Molecular Electronic-Structure
  Theory}   (John Wiley and Sons, Chichester, UK, 2000).

\bibitem{Sherrill2010}
C. Sherrill, J. Chem. Phys. \textbf{132}, 110902 (2010).

\bibitem{LJL+10}
T.D. Ladd, F. Jelezko, R. Laflamme, Y. Nakamura, C. Monroe and J.L. O'Brien,
  Nature \textbf{464}, 45 (2010).

\bibitem{Fey82}
R. Feynman, Optics News (now OPN) \textbf{11} (11), 11--22 (1982).

\bibitem{Llo96}
S. Lloyd, Science \textbf{273}, 1073 (1996).

\bibitem{AGDL+05}
A. Aspuru-Guzik, A. Dutoi, P. Love and M. Head-Gordon, Science \textbf{309},
  1704 (2005).

\bibitem{LW99}
D. Lidar and H. Wang, Phys. Rev. E \textbf{59}, 2429 (2008).

\bibitem{KJLM+08}
I. Kassal, S.P. Jordan, P.J. Love, M. Mohseni and A. Aspuru-Guzik, Proc. Natl.
  Acad. Sci. \textbf{105}, 18681 (2008).

\bibitem{OGKL01}
G. Ortiz, J. Gubernatis, E. Knill and R. Laflamme, Phys. Rev. A \textbf{64},
  022319 (2001).

\bibitem{KA09}
I. Kassal and A. Aspuru-Guzik, J. Chem. Phys. \textbf{131}, 224102 (2009).

\bibitem{Kassal2010}
I. Kassal, J. Whitfield, A. Perdomo-Ortiz, M.H. Yung and A. Aspuru-Guzik,
  preprint, arXiv:1007.2648 (2010).  $<${http://arxiv.org/abs/1007.2648}$>$.

\bibitem{NC01}
M. Nielsen and I. Chuang, \emph{Quantum Computation and Quantum Information}
  (Cambridge University Press, Cambridge, UK, 2001).

\bibitem{Got09}
D. Gottesman, preprint, arXiv:0904.2557 (2009).
  $<${http://arxiv.org/abs/0904.2557}$>$.

\bibitem{Kit97}
A.Y. Kitaev, Russian Math. Surveys \textbf{52}, 1191 (1997).

\bibitem{CBMG08}
C.R. Clark, T.S. Metodi, S.D. Gasster and K.R. Brown, Phys. Rev. A \textbf{79},
  062314 (2009).

\bibitem{LW+09}
B.P. Lanyon, J.D. Whitfield, G.G. Gillet, M.E. Goggin, M.P. Almeida, I. Kassal,
  J.D. Biamonte, M. Mohseni, B.J. Powell, M. Barbieri, A. Aspuru-Guzik and A.G.
  White, Nature Chem. \textbf{2}, 106 (2010).

\bibitem{DXP+09}
J. Du, N. Xu, X. Peng, P. Wang, S. Wu and D. Lu, Phys. Rev. Lett \textbf{104},
  030502 (2010).

\bibitem{Zwanzig1965}
R. Zwanzig, Time-Correlation Functions and Transport Coefficients in
  Statistical MechanicsAnn. Rev. Phys. Chem. \textbf{16}, 67--102 (1965).

\bibitem{Heller1981}
E. Heller, The Semiclassical Way to Molecular SpectroscopyAccounts of Chemical
  Research, Acc. Chem. Res. \textbf{14}, 368--375 (1981).

\bibitem{Feit1982}
M. Feit, J. J.A.~Fleck and A. Steiger, Journal of Computational Physics, J.
  Comp. Phys. \textbf{47}, 412--433 (1982).

\bibitem{Ceotto2009}
S.A. M.~Ceotto, G.F. Tantardini and A. Aspuru-Guzik, Multiple coherent states
  for first-principles semiclassical initial value representation molecular
  dynamicsThe Journal Chemical Physics, J. Chem. Phys. \textbf{130}, 234113
  (2009).

\bibitem{Fel96}
D. Feller, J. Comp. Chem. \textbf{17}, 1571 (1996).

\bibitem{SDES+07}
K.L. Schuchardt, B.T. Didier, T. Elsethagen, L. Sun, V. Gurumoorthi, J. Chase,
  J. Li and T.L. Windus, J. Chem. Inf. Model. \textbf{47}, 1042 (2007).

\bibitem{DiV00}
D.P. DiVincenzo, Fortschr. Phys. \textbf{48}, 771 (2000), also
  arXiv:quant-ph/0002077.

\bibitem{JW28}
P. Jordan and E. Wigner, Z. Phys. A. \textbf{47}, 631 (1928).

\bibitem{HS05}
N. Hatano and M. Suzuki, in \emph{Quantum Annealing and Other Optimization
  Methods}, edited by A. Das and B.K. Chakrabarti  (Springer, Berlin, Germany,
  2005), Lectures Notes in Physics, pp. 37--68.

\bibitem{BACS06}
D.W. Berry, G. Ahokas, R. Cleve and B.C. Sanders, Commun. Math. Phys.
  \textbf{270}, 359--371 (2007).

\bibitem{Beals1998}
R. Beals, H. Buhrman, R. Cleve, M. Mosca and R. de~Wolf, Quantum Lower Bounds
  by Polynomials. in \emph{Proceedings of FOCS' 98}.  $<${also see,
  http://arxiv.org/abs/quant-ph/9802049}$>$, pp. 352--361.

\bibitem{SOGK+02}
R. Somma, G. Ortiz, J.E. Gubernatis, E. Knill and R. Laflamme, Phys. Rev. A
  \textbf{65}, 042323 (2002).

\bibitem{OHJ07}
E. Ovrum and M. Hjorth-Jensen, preprint, arXiv:0705.1928 (2007).
  $<${http://arxiv.org/abs/0705.1928}$>$.

\bibitem{PP00}
S. Parker and M. Plenio, Phys. Rev. Lett. \textbf{85}, 3049--52 (2000).

\bibitem{Kit95}
A. Kitaev, preprint, arXiv:quant-ph/9511026 (1995).
  $<${http://arxiv.org/abs/quant-ph/9511026}$>$.

\bibitem{TN06}
A. Tomita and K. Nakamura, Int. J. Quantum Information \textbf{2}, 119 (2004).

\bibitem{DJSW06}
M. Dob\v{s}\'{i}\v{c}ek, G. Johansson, V. Shumeiko and G. Wendin, Phys. Rev. A
  \textbf{76}, 030306(R) (2007).

\bibitem{XMJ+07}
L. Xiu-Mei, L. Jun and S. Xian-Ping, Chinese Phys. Lett. \textbf{24}, 3316
  (2007).

\bibitem{BCC06}
K.R. Brown, R.J. Clark and I.L. Chuang, Phys. Rev. Lett. \textbf{97}, 050504
  (2006).

\bibitem{FGGS00}
E. Farhi, J. Goldstone, S. Gutmann and M. Sipser, Science \textbf{292}, 472
  (2000).

\bibitem{WBL02}
L.A. Wu, M.S. Byrd and D.A. Lidar, Phys. Rev. Lett. \textbf{89}, 057904 (2002).

\bibitem{PVAA+08}
A. Perdomo-Ortiz, S.E. Venegas-Andraca and A. Aspuru-Guzik, Quantum Information
  Processing pp. 1--20 (2010).

\bibitem{BKS09}
S. Boixo, E. Knill and R.D. Somma, Quant. Inf. Comp. \textbf{9}, 0833 (2009).

\bibitem{WA08}
P. Wocjan and A. Abeyesinghe, Phys. Rev. A \textbf{78}, 042336 (2008).

\bibitem{KKR06}
J. Kempe, A. Kitaev and O. Regev, SIAM J. Computing \textbf{35} (5), 1070--1097
  (2006).

\bibitem{OT06}
R. Oliveira and B. Terhal, Quant. Inf. Comp. \textbf{8}, 0900 (2008).

\bibitem{WKAG+08}
H. Wang, S. Kais, A. Aspuru-Guzik and M. Hoffmann, Phys. Chem. Chem. Phys.
  \textbf{10}, 5388 (2008).

\bibitem{WAF09}
H. Wang, S. Ashhab and F. Nori, Phys. Rev. A \textbf{79}, 042335 (2009).

\bibitem{WKA09}
N. Ward, I. Kassal and A. Aspuru-Guzik, J. Chem. Phys. \textbf{130}, 194105
  (2009).

\bibitem{Veis2010}
L. Veis and J. Pittner, preprint, arXiv:1008.3451 (2010).
  $<${http://arxiv.org/abs/1008.3451}$>$.

\bibitem{PyQ}
R.P. Muller, Python Quantum Chemistry (PyQuante) program, version 1.6 (Sandia
  National Laboratories,  Albuquerque, NM,  2007).

\end{thebibliography}
\newpage
\renewcommand{\thetable}{A\arabic{table}}
\setcounter{table}{0}
\begin{table}[ht]
    \tbl{
    {The quantum circuits corresponding to evolution of the listed Hermitian second-quantized operators}. Here, $p$, $q$, $r$, and $s$ are orbital indices corresponding to qubits such that qubit state $\ket{1}$ indicates an occupied orbital and $\ket{0}$ indicates unoccupied. It is assumed that the orbital indices satisfy $p>q>r>s$. These circuits were found by performing the Jordan-Wigner transformation given in Eqs.~\eqref{subeq:JW(crea)} and \eqref{subeq:JW(dest)} and then propagating the obtained Pauli spin variables~\cite{SOGK+02}. In each circuit, $\theta=\theta(h)$ where $h$ is the integral preceding the operator.  Gate ${\sf T}(\theta)$ is defined by ${\sf T}\ket{0}=\ket{0}$ and ${\sf T}\ket{1}=\exp(-i\theta)\ket{1}$, ${\sf G}$ is the global phase gate given by $\exp(-i\phi)\mathbf{1}$, and the change-of-basis gate ${\sf Y}$ is defined as ${R}_x(-\pi/2)$. Gate ${\sf H}$ refers to the Hadamard gate.  For the number-excitation operator, both ${\sf M}={\sf Y}$ and ${\sf M}={\sf H}$ must be implemented in succession. Similarly, for the double excitation operator each of the 8 quadruplets must be implemented in succession.   The global phase gate must be included due to the phase-estimation procedure.  Phase estimation requires controlled versions of these operators which can be accomplished by changing all gates with $\theta$-dependence into controlled gates dependent on register qubits. }
    { \begingroup
    \everymath{\scriptstyle}
    \begin{tabular}{lll}
 \toprule
 \multicolumn{2}{l}{Second quantized operators} ~ ~ ~ ~ ~ ~&\multicolumn{1}{l}{Circuit}\\
 \colrule
         \begin{tabular}{l}Number\\ operator\end{tabular}
            &$h_{pp}\ad_p a_p$
            &$\Qcircuit@C=.3em @R.1em @!R{&\qw&\gate{{\sf T}(\theta)}&\qw&\qw}$\\
                 ~&~&~\\
         \begin{tabular}{l}Excitation\\operator\end{tabular}
            &$h_{pq}(\ad_p a_q +  \ad_q a_p)$
            & $\Qcircuit @C=.5em @R=1.1em @!R {
                 \lstick{p}&\qw&\gate{{\sf H}}\qw      &\ctrldt{2}     &\qw               & \ctrldt{2}& \gate{{\sf H}}&\gate{{\sf Y}}     & \ctrldt{2}    & \qw           &\ctrldt{2}      &\gate{{\sf Y}}&\qw  \\
		           &/\qw   & \qw               &\qw            &\qw                & \qw      &  \qw           &\qw               &\qw             &\qw           &\qw            &\qw &\qw\\ 
                 \lstick{q}&\qw&\gate{{\sf H}}\qw      &\targ          & \gate{R_z(\theta)}& \targ     & \gate{{\sf H}}&\gate{{\sf Y}}     &\targ          &\gate{R_z(\theta)}      &\targ          &\gate{{\sf Y}}&\qw}$ \\
                 ~&~&~\\
         \begin{tabular}{l}Coulomb and  \\exchange operators \end{tabular}
            &$h_{pqqp} \ad_p \ad_q a_q a_p$
            & $\Qcircuit@C=.3em @R.1em @!R{
 \lstick{p}&\qw  &\gate{{\sf G}(\theta)}   &\gate{R_z(\theta)}&\ctrl{2}   &\qw        &\ctrl{2}&\qw\\
            &/\qw   & \qw                &\qw             &\qw            & \qw      &  \qw   &\qw   \\
 \lstick{q}&\qw  &\qw             &\gate{R_z(\theta)}&\targ      &\gate{R_z(\theta)} &\targ&\qw
            }$ \\
                 \begin{tabular}{l}
			 Number-excitation$^{\rm a}$\\
                operator
            \end{tabular}&
            \begin{tabular}{cl}
                $h_{pqqr}$
                &$(\ad_p \ad_q a_q a_r $\\
                & $+ \ad_r \ad_q a_q a_p)$
            \end{tabular}&
            \multicolumn{1}{c}{
 $\Qcircuit@C=.3em @R.1em @!R{
 \lstick{p}  &\qw&\gate{\sf M}&\ctrldt{2}    &\qw            & \qw           & \qw           & \qw         &\qw      &\ctrldt{2} &\gate{\sf M}   & \qw               \\
 	     &/\qw&\qw        &\qw          &   \qw          &\qw            &\qw            &\qw          &\qw      &\qw       &\qw            &\qw  \\
 \lstick{q+1}&\qw&\qw        &\targ         &\ctrl{2}       &\qw            &\qw            &\qw          &\ctrl{2} &\targ      &\qw            &\qw  \\
 \lstick{q}  &\qw&\qw        &\qw           & \qw           & \qw           &\qw            & \qw         & \qw     &\qw        &\qw            &\qw                 \\
 \lstick{q-1}&\qw&\qw         & \qw         &\targ          & \ctrldt{2}    & \qw           & \ctrldt{2}  &\targ    & \qw       &\qw            &\qw \\
 	     &/\qw&\qw        &\qw          &   \qw          &\qw            &\qw            &\qw          &\qw      &\qw       &\qw            &\qw  \\
 \lstick{r}  &\qw&\gate{\sf M}& \qw     &\qw            & \targ         & \gate{R_z(\theta)}    & \targ       &\qw      & \qw       &\gate{\sf M}   &\qw
 }$  \begin{tabular}{l}~\\~\\where ${\sf M}= \{{\sf H},{\sf Y}\}$\end{tabular}}\\
        \begin{tabular}{l}Double excitation\\
            operator
        \end{tabular}&
        \begin{tabular}{cl}
                $h_{pqrs}$  &   $(\ad_p \ad_q a_r a_s$\\
                        & $+\ad_s \ad_r a_q a_p)$
        \end{tabular}
            &
 $\Qcircuit@C=.3em @R.1em @!R{
 \lstick{p}              & \qw & \gate{\sf M_1}      &\ctrldt{2}&\qw     & \qw               & \qw                   &\qw     &\qw     &\ctrldt{2}&\gate{\sf M_1^\dagger}&\qw\\
 & \qw &/\qw             &\qw      &\qw      &\qw                &\qw                    &\qw     &\qw     &\qw     &\qw               &\qw\\
 \lstick{q}& \qw & \gate{\sf M_2}            & \targ   & \ctrl{2}& \qw               &\qw                    &\qw     &\ctrl{2}&\targ   &\gate{\sf M_2^\dagger}&\qw\\
 & \qw &/\qw             &\qw      &\qw      &\qw                &\qw                    &\qw     &\qw     &\qw     &\qw               &\qw\\
 \lstick{r}& \qw & \gate{\sf M_3}    & \qw     & \targ   & \ctrldt{2}        &\qw                    &\ctrldt{2}&\targ &\qw     &\gate{\sf M_3^\dagger}&\qw\\
 & \qw &/\qw             &\qw      &\qw      &\qw                &\qw                    &\qw     &\qw     &\qw     &\qw               &\qw\\
 \lstick{s}& \qw & \gate{\sf M_4}    & \qw     &\qw      & \targ             & \gate{R_z(\theta)}    &\targ   &\qw     &\qw     &\gate{\sf M_4^\dagger}&\qw
 }$ \begin{tabular}{l}~\\  where $\sf ( M_1,M_2,M_3,M_4)$=\\
                 \{$({\sf H},{\sf H},{\sf H},{\sf H})$, $({\sf Y},{\sf Y},{\sf Y},{\sf Y})$,\\
                 $({\sf H},{\sf Y},{\sf H},{\sf Y})$,$({\sf Y},{\sf H},{\sf Y},{\sf H})$,\\
                 $({\sf Y},{\sf Y},{\sf H},{\sf H})$,$({\sf H},{\sf H},{\sf Y},{\sf Y})$, \\
                 $({\sf Y},{\sf H},{\sf H},{\sf Y})$, $({\sf H},{\sf Y},{\sf Y},{\sf H})$\} \end{tabular}\\
 \botrule
 \toprule
 &Notation:& $\Qcircuit @C=.5em @R=0em @!R {
 & \qw &\ctrldt{3}       &\qw            &\ctrldt{3}& \qw    &~&~& &~&~&~& \qw &\ctrl{1} &\qw            & \qw           & \qw            &\qw   &\qw      &\ctrl{1}&\qw\\
 & \qw &\qw              &\qw            &\qw       & \qw    &~&~& &~&~&~& \qw &\targ    &\ctrl{1}       &\qw            &\qw             &\qw   &\ctrl{1} &\targ   &\qw\\
 & \qw & \qw             &\qw            &\qw       & \qw    &~&~&\equiv&~&~&~& \qw &\qw         &\targ          & \ctrl{1}      & \qw            &\ctrl{1}&\targ          &\qw  &\qw\\
 & \qw &\targ            &\gate{~}       &\targ     & \qw    &~&~& &~&~&~& \qw &\qw      &\qw            &\targ          &\gate{~}        &\targ &\qw     &\qw     &\qw
 }$\\
         \botrule
 \end{tabular}
 \endgroup}
\tabnote{$^{\rm a}$The spin variable representation of this operator depends on whether $q$ lies in the range $p$ to $r$ or outside of it.}
\label{tblf:cir_rep}
\end{table}

\begin{sidewaystable}[ht]
	\tbl{The spin variable representation of second quantized Hermitian operators after performing the Jordan-Wigner transformation to yield tensor products of Pauli sigma matrices (spin variables) that have the proper anti-commutation relations. See Eqs.~\eqref{subeq:JW(crea)} and \eqref{subeq:JW(dest)} for the form of the transformation used.  The subscripts label the qubit and the molecular spin orbital that corresponds to that qubit.  The pre-factors $h_{ij}$ and $h_{ijkl}$ are one- and two- electron integrals given in Eqs.~\eqref{eq:1eint} and \eqref{eq:2eint}.  In our algorithm, these are calculated using a classical computer.  Rarely is this the limiting step since the basis sets are typically chosen for ease of integration.  In this table, we provide for the case that the one- and two-electron integrals are complex.  The integrals $h_{pqqp}$ and $h_{pqpq}$ are referred to as the Coulomb and exchange integrals, respectively. The exchange operator has the same representation (with opposite sign) as the Coulomb integral due to the commutation relations.  The circuit representation of the exponentiation of each of these term is given in {Table}~\ref{tblf:cir_rep}. }
	{\small\begin{tabular}{llll}
\toprule
Description&Second Quantization$^{\rm a}$&\multicolumn{2}{l}{Pauli representation}\\
\colrule
        \begin{tabular}{l}Number\\ Operator\end{tabular} &$h_{pp}\ad_p a_p$                &\multicolumn{2}{l}{$\frac{h_{pp}}{2}(\I_p-\sigma^{z}_p)$}\\
        \begin{tabular}{l}Excitation\\Operator\end{tabular}&$h_{pq}\ad_p a_q + h_{qp}  \ad_q a_p$ &\multicolumn{2}{l}{${\displaystyle \frac{1}{2}\left(\bigotimes_{k=q+1}^{p-1}\sigma^{z}_k\right)}\left(\begin{tabular}{l}$\Re\{h_{pq}\}(\sigma^{x}_q\sigma^{x}_p+\sigma^{y}_q\sigma^{y}_p)$\\$+\Im\{h_{pq}\}(\sigma^{y}_q\sigma^{x}_p-\sigma^{x}_q\sigma^{y}_p)$   \end{tabular}\right)$}\\
        \begin{tabular}{l}Coulomb \\Operators \end{tabular}&$h_{pqqp} \ad_p \ad_q a_q a_p$       & \multicolumn{2}{l}{${\displaystyle\frac{h_{pqqp}}{4}\left(\I-\sigma^{z}_p-\sigma^{z}_q+\sigma^{z}_p \sigma^{z}_q\right)} $}\\
		\begin{tabular}{l}Number with$^{\rm b}$\\ Excitation Operator\end{tabular}&
            $h_{pqqr} \ad_p \ad_q a_q a_r +h_{rqqp} \ad_r \ad_q a_q a_p$
            &\multicolumn{2}{l}{${
            \displaystyle
            \left( \bigotimes_{k=r+1}^{p-1}\sigma^{z}_k\right)

            \left[
                 \left(
            \begin{tabular}{l}
                $\Re\{h_{pqqr}\}(\sigma^{x}_r\sigma^{x}_p+\sigma^{y}_r\sigma^{y}_p)$\\
                $+\Im\{h_{pqqr}\}(\sigma^{y}_r\sigma^{x}_p-\sigma^{x}_r\sigma^{y}_p)$
            \end{tabular}
                \right)
            -\sigma^{z}_q\left(
            \begin{tabular}{l}
                $\Re\{h_{pqqr}\}(\sigma^{x}_r\sigma^{x}_p+\sigma^{y}_r\sigma^{y}_p)$\\
                $+\Im\{h_{pqqr}\}(\sigma^{y}_r\sigma^{x}_p-\sigma^{x}_r\sigma^{y}_p)$
            \end{tabular}
                \right)
            \right]
            }
            $ }\\
        \begin{tabular}{l}Double\\Excitation\\Operator\end{tabular}&$h_{pqrs} \ad_p \ad_q a_r a_s+h_{srqp} \ad_s \ad_r a_q a_p$&\
            $\displaystyle \left(\bigotimes_{k=s+1}^{r-1}\sigma^{z}_k\right)\left(\bigotimes_{k=q+1}^{p-1}\sigma^{z}_k\right)$\
            &$\left(\begin{tabular}{l}
                $\frac{\Re\{h_{pqrs}\}}{8}\left(
                        \begin{tabular}{rl}&$\sigma^{x}_s\sigma^{x}_r\sigma^{x}_q \sigma^{x}_p-\sigma^{x}_s\sigma^{x}_r\sigma^{y}_q\sigma^{y}_p+\sigma^{x}_s\sigma^{y}_r\sigma^{x}_q\sigma^{y}_p$\\
                                      $+$&$\sigma^{y}_s\sigma^{x}_r\sigma^{x}_q\sigma^{y}_p+\sigma^{y}_s\sigma^{x}_r\sigma^{y}_q\sigma^{x}_p-\sigma^{y}_s\sigma^{y}_r\sigma^{x}_q\sigma^{x}_p$\\
                                  $+$&$\sigma^{x}_s\sigma^{y}_r\sigma^{y}_q\sigma^{x}_p+\sigma^{y}_s\sigma^{y}_r\sigma^{y}_q\sigma^{y}_p$
                        \end{tabular}\right)$\\
                        +$\frac{\Im\{h_{pqrs}\}}{8}\left(
                        \begin{tabular}{rl}&$\sigma^{y}_s\sigma^{x}_r\sigma^{x}_q\sigma^{x}_p+\sigma^{x}_s\sigma^{y}_r\sigma^{x}_q\sigma^{x}_p-\sigma^{x}_s\sigma^{x}_r\sigma^{y}_q\sigma^{x}_p$\\
                                      $-$&$\sigma^{x}_s\sigma^{y}_r\sigma^{y}_q\sigma^{y}_p-\sigma^{y}_s\sigma^{x}_r\sigma^{y}_q\sigma^{y}_p+\sigma^{y}_s\sigma^{y}_r\sigma^{x}_q\sigma^{y}_p$\\
                                  $+$&$\sigma^{y}_s\sigma^{y}_r\sigma^{x}_q\sigma^{y}_p+\sigma^{y}_s\sigma^{y}_r\sigma^{y}_q\sigma^{x}_p$
                        \end{tabular}\right)$
             \end{tabular}\right)$\\
        \botrule
    \end{tabular}}
    \tabnote{$^{\rm a}$It is assumed that $p>q>r>s$ for all cases listed.}
\tabnote{$^{\rm b}$The spin variable representation of this operator depends if $q$ is an orbital in the range $p$ to $r$ or if it is outside this range.}
\label{tbl:spin_vars}
\end{sidewaystable}

\begin{sidewaystable}[t]
\tbl{A circuit that implements the unitary propagator associated with the two electron operators of the H$_2$ Hamiltonian in the minimal basis.
		The circuit is derived from the two non-vanishing real integrals of the two-electron interaction and the corresponding operators.  In the circuit,
		$\theta\equiv h_{1423}+h_{1243}=.36257$.  In Eq.~\ref{eq:EEop} two of the four pre-factors vanish due to the spin orthogonality however, in general, there would be eight terms to simulate.}
{$
\Qcircuit @C=.4em @R=.5em @!R {
\lstick{PEA}   &\qw		&\qw		&\qw		&\qw		&\ctrl{4}		&\qw		&\qw		&\qw		&\qw	  	        &\qw		&\qw		&\qw		&\qw		&\ctrl{4}	        	&\qw		&\qw		&\qw		&\qw		        &\qw		&\qw		&\qw		&\qw		&\ctrl{4}		&\qw		&\qw		&\qw					&\qw		&\qw		&\qw		&\qw		&\qw		&\ctrl{4}		&\qw		&\qw		&\qw		&\qw		\\
\lstick{\chi_1}&\gate{\sf H}	&\ctrl{1}	&\qw		&\qw		&\qw			&\qw		&\qw		&\ctrl{1}	&\gate{\sf H}	        &\gate{\sf Y}	&\ctrl{1}	&\qw		&\qw		&\qw		        	&\qw		&\qw		&\ctrl{1}	&\gate{\sf Y^\dagger}	&\gate{\sf H}	&\ctrl{1}	&\qw		&\qw		&\qw			&\qw		&\qw		&\ctrl{1}				&\gate{\sf H}	        &\gate{\sf Y}	&\ctrl{1}	&\qw		&\qw		&\qw			&\qw		&\qw		&\ctrl{1}	&\gate{\sf Y^\dagger}	\\
\lstick{\chi_2}&\gate{\sf H}	&\targ		&\ctrl{1}	&\qw		&\qw			&\qw		&\ctrl{1}	&\targ		&\gate{\sf H}	        &\gate{\sf Y}	&\targ		&\ctrl{1}	&\qw		&\qw	        		&\qw		&\ctrl{1}	&\targ		&\gate{\sf Y^\dagger}	&\gate{\sf Y}	&\targ		&\ctrl{1}	&\qw		&\qw			&\qw		&\ctrl{1}	&\targ					&\gate{\sf Y^\dagger}	&\gate{\sf H}	&\targ		&\ctrl{1}	&\qw		&\qw			&\qw		&\ctrl{1}	&\targ		&\gate{\sf H}	\\
\lstick{\chi_3}&\gate{\sf Y}	&\qw		&\targ		&\ctrl{1}	&\qw			&\ctrl{1}	&\targ		&\qw		&\gate{\sf Y^\dagger}	&\gate{\sf H}	&\qw		&\targ		&\ctrl{1}	&\qw			        &\ctrl{1}	&\targ		&\qw		&\gate{\sf H}    	&\gate{\sf Y}	&\qw		&\targ		&\ctrl{1}	&\qw			&\ctrl{1}	&\targ		&\qw					&\gate{\sf Y^\dagger}	&\gate{\sf H}	&\qw		&\targ		&\ctrl{1}	&\qw			&\ctrl{1}	&\targ		&\qw		&\gate{\sf H}	\\
\lstick{\chi_4}&\gate{\sf Y}	&\qw		&\qw		&\targ		&\gate{R_z(-\frac{\theta}{4}t)}	&\targ		&\qw		&\qw	&\gate{\sf Y^\dagger}	&\gate{\sf H}	&\qw		&\qw		&\targ		&\gate{R_z(-\frac{\theta}{4}t)}	&\targ		&\qw		&\qw		&\gate{\sf H}	        &\gate{\sf H}	&\qw		&\qw		&\targ		&\gate{R_z(\frac{\theta}{4}t)}	&\targ		&\qw		&\qw		                &\gate{\sf H}	        &\gate{\sf Y}	&\qw		&\qw		&\targ		&\gate{R_z(\frac{\theta}{4}t)}	&\targ		&\qw		&\qw		&\gate{\sf Y^\dagger}	
	}$}
	\label{cir:4func}
\end{sidewaystable}

\end{document}